\documentclass{aa}
\usepackage[varg]{txfonts}
\bibpunct{(}{)}{;}{a}{}{,}

\usepackage{amsmath,amssymb}
\usepackage{mathtext}
\usepackage[T1,T2A]{fontenc}
\usepackage[utf8]{inputenc}
\usepackage[english]{babel}
\usepackage{soul}
\usepackage{float}
\usepackage{caption} 

\begin{document}

\title{Investigation of gravitational stability of protoplanetary disks based on statistical analysis of their masses}

\titlerunning{Gravitational stability of PPDs}

\author{Sophia~A.~Drobchik \and Sergey~A.~Khaibrakhmanov}

\authorrunning{S.~A.~Drobchik and S.~A.~Khaibrakhmanov}

\institute{
  Saint Petersburg State University, Saint Petersburg, Russia\\
  \email{s.a.drobchik@gmail.com}  
}


\abstract{
We compiled a sample of $1155$ protoplanetary disks, combining data from ten surveys of star-forming regions. Based on the sample, we constructed a power-law approximation of the disk mass distribution: $dN/dM \propto M^{-\beta}$, $\beta = 1.36 \pm 0.14$. We used the sample for a statistical analysis of the gravitational stability of protoplanetary disks. To analyze the stability of the disks, we calculated the Toomre parameter ($Q$) for each of them. In the calculations, it was assumed that the radial density distribution in the disks is described by a power-law profile. The calculations of the Toomre parameter show that only $1.2$~\% of the disks in the sample are formally unstable ($Q < 1$), while $1.7$~\% are in a state of marginal stability ($1 \leq Q \leq 2$). The low observed abundance of unstable disks contradicts theoretical expectations and may be explained by a systematic underestimation of disk masses due to limitations of observational methods. Considering the effects of optical depth, CO depletion, as well as uncertainty in the gas-to-dust ratio, we conclude that the actual fraction of the gravitationally unstable systems may be significantly higher.
}


\maketitle

\nolinenumbers

\section{Introduction}
\label{sect:intro}

Accretion disks forming around young stars as a result of the gravitational collapse of rotating, magnetized molecular cloud cores exhibit diverse morphological features in observations with modern radio interferometric arrays such as ALMA (Atacama Large Millimeter/submillimeter Array) and VLA (Very Large Array). The observed structures include concentric rings, spiral arms, asymmetric structures possibly associated with vortices, as well as so-called `gaps` (ring-like regions of reduced brightness) and `cavities` (essentially non-emitting annular regions). The presence of such substructures may be linked to planet formation processes \citep{Keppler2018} and allows us to interpret these systems as protoplanetary disks (PPDs). To analyze the transition from accretion disks to protoplanetary disks, it is necessary to identify the physical conditions under which efficient planet formation begins in the disks.

The primary mechanisms of planet formation are core accretion, whereby dust particles gradually coalesce to form planetary embryos, and gravitational instability, in which disk fragmentation leads to the formation of bound clumps under self-gravity (see, e.g., \citep{Kratter_2016, Armitage2025}). Gravitational instability develops when the balance between self-gravity and the stabilizing effects of gas pressure and rotation is disrupted. Such conditions usually arise in the outer regions of the disk ($> 50$--$100$~au), where low temperatures ($< 30$~K) and slow rotation create favorable conditions for its development.

Different planet formation mechanisms may dominate in various protoplanetary systems. For instance, \citet{Nguyen2024} showed that the formation of super-Jupiters with masses exceeding $4 M_{\text{Jup}}$ is possible via the core accretion mechanism, provided that the protoplanetary disk has sufficient metallicity. In systems such as HL Tau, the formation of giant planet cores through dust accretion is impeded by rapid dust migration toward the disk center, making gravitational instability a potentially more efficient mechanism \citep{Ueda2025}.

According to theoretical studies, the development of gravitational instability should lead to the formation of characteristic substructures in PPDs. 
Classical analytical works \citep{Jeans1902, Safronov1960, Toomre1964} formulated the criterion for gravitational instability, which quantitatively expresses the balance between the compressing force of the disk's self-gravity and the stabilizing factors --- gas pressure and centrifugal force. Numerical models \citep{Boss1997} show that gravitational instability 
in protostellar disks can lead to the formation of massive gaseous clumps. 
Further simulations incorporating gas heating and cooling (see, e.g., \citep{Vorobyov2005, Vorobyov2006, Basu2004}) demonstrate that spiral arms in unstable disks can fragment to form dense clumps, which in turn generates a non-steady episodic accretion regime. Modern three-dimensional simulations \citep{Xu2025} indicate that spiral perturbations induced by gravitational 
instability in disks may explain the origin of substructures observed with ALMA.

The interpretation of observed substructures is complicated by the high optical depth of disks in millimeter waves and the limited angular resolution of telescopes. Assessing the potential for gravitational instability in such systems requires an analysis of fundamental disk parameters. Among these, mass is of particular interest, as it determines the disk's susceptibility to gravitational collapse and fragmentation.

Contemporary methods for mass estimation are subject to significant systematic uncertainties, leading to a potential underestimation of actual values. Determining the dust mass from millimeter continuum emission requires knowledge of several system parameters. Such characteristics as dust temperature, its opacity, and the particle size distribution cannot be precisely constrained from observations and are typically assigned based on standard theoretical models for PPDs \citep{Semenov2003}. In particular, the opacity value is influenced by the chemical composition, shape, and internal structure of dust grains, which may differ substantially from average model values, as confirmed by laboratory measurements \citep{Agladze1996}.

Estimating the gas mass, which constitutes the bulk of the protostellar system`s material, relies on observations of molecular lines, primarily CO and its isotopologues. However, these methods face two major challenges (see, e.g., the review \citet{Williams2011}). First, CO lines are often optically thick in dense disk regions, complicating accurate determination of the total molecular gas content \citep{vanZadelhoff2001, Thi2001}. Second, chemical depletion of CO is observed --- its freeze-out onto dust grains in cold disk regions, as well as the conversion of CO into more complex molecules, making this molecule an unreliable tracer of the total gas mass \citep{Aikawa1996}. These effects result in extremely low estimates of gas masses and gas-to-dust ratios derived from observations of CO isotopologues \citep{Miotello2017}. Furthermore, a substantial fraction of the gas may exist in the form of molecular hydrogen H$_2$, which does not emit under PPD conditions.

The aforementioned factors --- the high optical depth of the material, the chemical depletion of CO, and the uncertainty in the gas-to-dust ratio --- lead to systematic underestimates of disk masses by an order of magnitude or more \citep{Miotello2017}. Accurate determination of the disk mass is crucial for understanding the dynamical evolution of protoplanetary disks, as it governs their stability against gravitational perturbations. Statistical analysis of observational data enables the estimation of the mass distribution and the identification of systems in which gravitational instability may develop. This approach allows us to assess the prevalence of conditions favorable to fragmentation, even in the presence of the systematic uncertainties described above.

Within the framework of statistical studies of protoplanetary disk properties, the work by \citet{Manara2023} analyzing a sample of $\sim 890$ objects is among the most extensive. The authors investigated the dependencies between dust mass, accretion rate, stellar mass, and other parameters. In that study, stellar masses for a significant fraction of objects were determined with low precision, and variations in the gas-to-dust ratio for objects of different evolutionary classes (from Class 0 to Class III) were not taken into account, although this ratio may differ substantially across evolutionary stages.

The aim our work is to perform a statistical analysis of a large sample of protoplanetary disks in order to identify systems where conditions are favorable for the development of gravitational instability. Our main task is to analyze the Toomre parameter ($Q$) as an indicator of stability and to study its dependence on other system characteristics.

Section~\ref{sect:methods} describes the theoretical approach to analyzing gravitational stability based on the Toomre criterion and provides the formulas for calculating it from observable system parameters. Section~\ref{sect:sample} outlines the methodology for constructing a sample of objects with reliably measured physical parameters using data from ten major astronomical catalogs. In Section~\ref{sect:results}, the Toomre parameter is computed for each disk using the sample data, a statistical analysis of its distribution is performed, and its correlations with key system characteristics (such as disk mass and stellar mass) are investigated, along with a classification of the disks according to their stability status. 
Finally, Section~\ref{sect:outro} summarizes the results of this analysis: the fraction of protoplanetary disks in the sample that are potentially susceptible to gravitational instability is determined, and the findings are discussed in the context of theoretical predictions and observed disk morphology.

\section{Methods for Analyzing Gravitational Stability}
\label{sect:methods}

\subsection{Gravitational Stability Criterion}

The gravitational stability of protoplanetary disks is usually analyzed using the method of small perturbations. One consider the evolution of surface density perturbations in a thin, self-gravitating disk that rotates in the gravitational field of the central star. The Toomre parameter $Q$ quantifies the balance between the destabilizing effect of self-gravity and the stabilizing effects of gas pressure and rotation \citep{Safronov1960, Toomre1964}:

\begin{equation}
    Q = \frac{c_s \kappa}{\pi G \Sigma},
\end{equation}
where $c_s$ is the sound speed in the medium, $\kappa$ is the epicyclic frequency, $G$ is the gravitational constant, and $\Sigma$ is the surface density of the disk. The criterion for gravitational stability with respect to local axisymmetric perturbations is formulated in terms of this parameter: the disk is gravitationally stable if $Q > 1$. Violation of this condition ($Q < 1$) indicates the development of gravitational instability, which can lead to disk fragmentation and the formation of clumps.

Estimating the Toomre parameter $Q$ allows disks to be classified as gravitationally unstable ($Q < 1$), marginally stable ($1 \leq Q \leq 2$) \citep{Kratter_2016}, and gravitationally stable ($Q > 2$).

\subsection{Disk Model and Calculation Parameters}
\label{sect:model}

To estimate the Toomre parameter $Q$, one need the values of the surface density $\Sigma$, sound speed $c_s$, and epicyclic frequency $\kappa$. We calculate these quantities using a disk model with a power-law surface density profile, typical of PPDs \citep{Williams2011}. In this approximation, the surface density decreases with radius as:

\begin{equation}
    \Sigma(r) = \Sigma_0 \left(\frac{r}{r_0}\right)^{-1},
    \label{eq:S(r)}
\end{equation}
where $r_0$ is the characteristic disk radius and $\Sigma_0$ is the surface density at the radial distance $r_0$.

The surface density $\Sigma_0$ can be expressed in terms of the total disk mass $M_{\text{disk}}$:

\begin{equation}
    \Sigma_0 = \frac{M_{\text{disk}}}{2\pi r_0^2},
\end{equation}
which follows from the normalization condition $\int_{R_{\text{in}}}^{R_{\text{out}}} \Sigma(r) \cdot 2\pi r dr = M_{\text{disk}}$, 
where $R_{\text{in}}$ and $R_{\text{out}}$ are the inner and outer radii of the disk. For the purposes of our statistical analysis, we adopt $R_{\text{out}} = r_0$, where $r_0$ is the characteristic radius used in the model.

For a Keplerian disk, the epicyclic frequency is approximately equal to the angular velocity: $\kappa \approx \Omega = \sqrt{G M_* / r^3}$. To estimate the $Q$ parameter, we also need to specify the temperature profile. In this simplified model, a constant temperature $T = 20$~K, typical of the outer regions of PPDs \citep{Williams2011}, is adopted for the calculations.

If we take into account that the temperature varies with distance according to a power law $T(r) = T_0 (r/r_0)^{-q}$ with an exponent $q \approx 0.5$, then the sound speed becomes radius-dependent: $c_s \propto T^{1/2} \propto r^{-q/2}$. In this case, the radial dependence of the Toomre parameter is modified: $Q \propto r^{(1-q)/2} M_{\text{disk}}^{-1} M_*^{1/2}$. For $q=0.5$, the power-law index for $r$ becomes $0.25$ instead of $0.5$ in the isothermal case. This refinement does not alter the qualitative conclusions of the study, since a consistent methodology is applied to all disks for statistical comparison.

For the analysis of observational data, it is convenient to express the Toomre parameter in terms of the typical radius and mass of the disk, as well as the stellar mass:

\begin{equation}
Q \approx 18 \left(\frac{r_0}{100\,\text{au}}\right)^{1/2} 
          \left(\frac{M_{\text{disk}}}{0.01\,M_\odot}\right)^{-1} 
          \left(\frac{M_*}{1\,M_\odot}\right)^{1/2}.
\end{equation}

This formula shows that  the disk is gravitationally stable, $Q \approx 18$, for typical parameter values ($r_0 = 100$ au, $M_{\text{disk}} = 0.01 M_\odot$, $M_* = M_\odot$ \citep{Weidenschilling1977}). The transition to an unstable state ($Q < 1$) requires substantially higher disk masses relative to the stellar mass.

The derived formulas, given the specified disk parameters, can be used to classify observed objects based on the gravitational stability criterion. In the present work, we apply this criterion to perform a statistical analysis of a large sample of protoplanetary disks and examine the distribution of disks across stability categories. We determine the fraction of disks potentially susceptible to gravitational instability and analyze the dependence of the Toomre parameter $Q$ on the disk-to-star mass ratio $M_\text{disk} / M_\text{star}$. To estimate the $Q$ parameter for disks with uncertain parameters, we adopt standard values based on observational data \citep{Williams2011}: a temperature $T = 20$~K, typical for the outer regions of protoplanetary disks; a characteristic radius $r_0 = 100$~au, for objects without direct measurements of linear sizes; and a gas-to-dust mass ratio of $100$.

\section{Sample Description}
\label{sect:sample}

For a statistical assessment of the gravitational stability of PPDs, we have compiled a sample of observational data covering various star-forming regions. 
In contrast to the previous study by \citet{Manara2023} with a sample of about $890$ objects, the present work collects information from ten major catalogs, nearly doubling the statistical sample size. This approach ensures representativeness in terms of stellar masses, system ages, and environmental parameters.

\subsection{Star-Forming Regions}

The sample includes objects from the following star-forming regions:

\begin{enumerate}
    \item \textbf{Orion molecular clouds} --- a set of nearby active star-forming regions ($\sim400$~pc) that host numerous young clusters, including the Orion Nebula Cluster (ONC), the OMC1 molecular cloud, the Lynds 1641 region, and the NGC 2024 cluster.
    
    \item \textbf{Lupus complex} --- a young ($\sim1$--$3$~Myr) nearby star-forming region ($150$--$200$~pc) with intermediate stellar density.
    
    \item \textbf{Ophiuchus L1688} --- a dense cloud core within the Ophiuchus molecular cloud ($\sim140$~pc) with an age of $\sim0.5$--$1$~Myr.
    
    \item \textbf{Serpens} --- an active star-forming region at a distance of $\sim430$~pc, where young stellar objects appear both as isolated sources and within dense clusters \citep{Anderson2022}.
    
    \item \textbf{Taurus--Auriga} --- a nearby large star-forming region ($\sim140$~pc) characterized by low stellar density and the absence of massive OB stars.
    
    \item \textbf{Chamaeleon I} --- a nearby molecular cloud ($\sim160$--$180$~pc) with moderate stellar density and an age of $\sim2$--$3$~Myr. The region contains a well-studied population of young stellar objects that host protoplanetary disks \citep{Testi2022}.
    
    \item \textbf{Corona Australis} --- a compact molecular cloud at a distance of $\sim150$~pc, containing both scattered and clustered populations of young stellar objects spanning a range of ages \citep{Testi2022}.
    
    \item \textbf{Upper Scorpius} --- an older subgroup ($\sim5$--$10$~Myr) of the Scorpius--Centaurus association at a distance of $\sim145$~pc.
\end{enumerate}

\subsection{Sample Characteristics}

The compiled sample covers a wide range of physical parameters. The masses of the central stars range from brown dwarfs ($0.01 M_\odot$) to intermediate-mass stars ($>2 M_\odot$, with a maximum of $12 M_\odot$). The system ages span the early stages (Class 0/I) and transition disks (Class II). The environment includes both isolated objects and dense clusters with background ultraviolet radiation.

To obtain the data we used the following surveys, taken from open astronomical catalogs available through the \texttt{VizieR} database \footnote{\url{https://vizier.cds.unistra.fr}}. The final table of sample object parameters, compiled in this work, is available in the Zenodo repository \citep{Drobchik2025}.

\begin{enumerate}
    \item \citet{Anderson2022} presented a survey of 302 young stellar objects in the Serpens region (16 Class I objects, 35 flat-spectrum sources, 235 Class II, and 16 Class III), including measurements of disk continuum flux at 1.3 mm and $^{12}$CO (J=2-1) line data.
    
    \item Observations of 101 protoplanetary disks in the Lynds 1641 region (Orion A cloud) \citep{L1641-survey}, which include measurements in 1.3 mm continuum and $^{12}$CO, $^{13}$CO, C$^{18}$O (J=2-1) line data. The survey detected 89 disks in continuum (including 20 transition disks) out of a total of 23 transition disks in the sample.
    
    \item A survey of the 1-3 Myr old protoplanetary disk population in the Lupus complex \citep{Lupus-survey}, where observations include 1.33 mm continuum and CO J=2-1 lines with a spatial resolution of $\sim0.25``$. The survey sensitivity is 0.30 mJy ($3\sigma$), corresponding to a dust mass detection limit of $\sim0.2 M_{\oplus}$ per disk.
    
    \item A study of the Class II/F disk population in L1688 (Ophiuchus) conducted by \citet{Testi2022}, which includes a comparison with other star-forming regions (Lupus, Chamaeleon I, Corona Australis, Taurus, Upper Scorpius) of various ages.
    
    \item A study of 85 millimeter-optically thin Class III disks  \citep{ClassIII-survey}. The average dust mass per disk in the sample is $0.29 \pm 0.19 M_{\oplus}$.
    
    \item A survey of protoplanetary disks around brown dwarfs \citep{BD-survey}, which includes 49 objects from four star-forming regions (Ophiuchus, Taurus, Lupus, Upper Scorpius) and investigates disk evolution around substellar objects.
    
    \item Observations of disks in the Orion Nebula Cluster (ONC) and the OMC1 molecular cloud with a resolution down to $0.030``$ at 0.85 mm \citep{ONC-survey}. The survey detected 127 sources, including 15 new ones. Of these, 72 sources are spatially resolved at 3 mm, with sizes ranging from $\sim$8 to 100 au. The sources are classified into 76 ONC disks and the remainder as envelope-embedded OMC1 disks. Both samples exhibit anomalously small disk sizes and an absence of large disks ($>75$ au) compared to other nearby star-forming regions.
    
    \item A study of 179 disks in the core of the young 0.5 Myr cluster NGC 2024 \citep{NGC2024-survey}. ALMA observations with $0.25``$ resolution ($\sim$100 au) detected 57 disks, most of which are unresolved. Two populations are identified: an `eastern` population (located east of the cluster center) with a detection frequency of $45\pm7\%$ and masses comparable to disks in isolated regions, and a `western` population with a frequency of $15\pm4\%$, indicating strong external photoevaporation effects.
    
    \item The largest survey of 873 protoplanetary disks identified by Spitzer in the L1641 and L1647 regions (Orion A) \citep{OrionA-survey}. The survey detected 502 disks (58\%) with a median dust mass of $2.2^{+0.2}_{-0.2} M_{\oplus}$. The disk mass distribution is similar to that of other nearby star-forming regions aged 1-3 Myr. On scales of $\sim$100 pc, the median disk masses vary by less than 50\%, suggesting similar initial conditions for formation and disk evolution rates across different parts of the cloud.
    
    \item A survey of 92 protoplanetary disks in the $\sigma$ Orionis cluster ($\sim$3-5 Myr) \citep{SigmaOri-survey}, which contains data on 37 disks detected in millimeter continuum, with 11 having dust masses $M_{\text{dust}} \gtrsim 10 M_{\oplus}$. Dust masses decrease with proximity to the central O9-type star.
\end{enumerate}

To select objects suitable for analyzing disk gravitational stability, we applied the following criteria: the availability of a measured dust mass from millimeter-wavelength observations, a known distance to the star-forming region, and, where possible, an estimated mass of the central star. These criteria ensure the robustness of the Toomre parameter estimation and disk classification.

\section{Results}
\label{sect:results}

\subsection{Statistical Analysis of the Sample}

We analyzed the gravitational stability of protoplanetary disks based on the Toomre parameter $Q$ for the initial sample of $1706$ objects. We found $Q>200$ for $551$ disks, which correspond to extremely low surface densities. Such values indicate either highly evolved disks with a dissipated gas component or a systematic underestimation of masses in observational data. We excluded these objects from further analysis because even with a possible underestimation of masses by one to two orders of magnitude---typical for modern measurement methods \citep{Williams2011}---the $Q$ values would shift into the $1 < Q < 2$ range corresponding to marginal stability. Thus, these systems cannot be classified as gravitationally unstable under any realistic corrections to their masses. The final sample contains $1155$ disks.

\begin{figure}[H]
    \centering
    \includegraphics[width=0.99\linewidth]{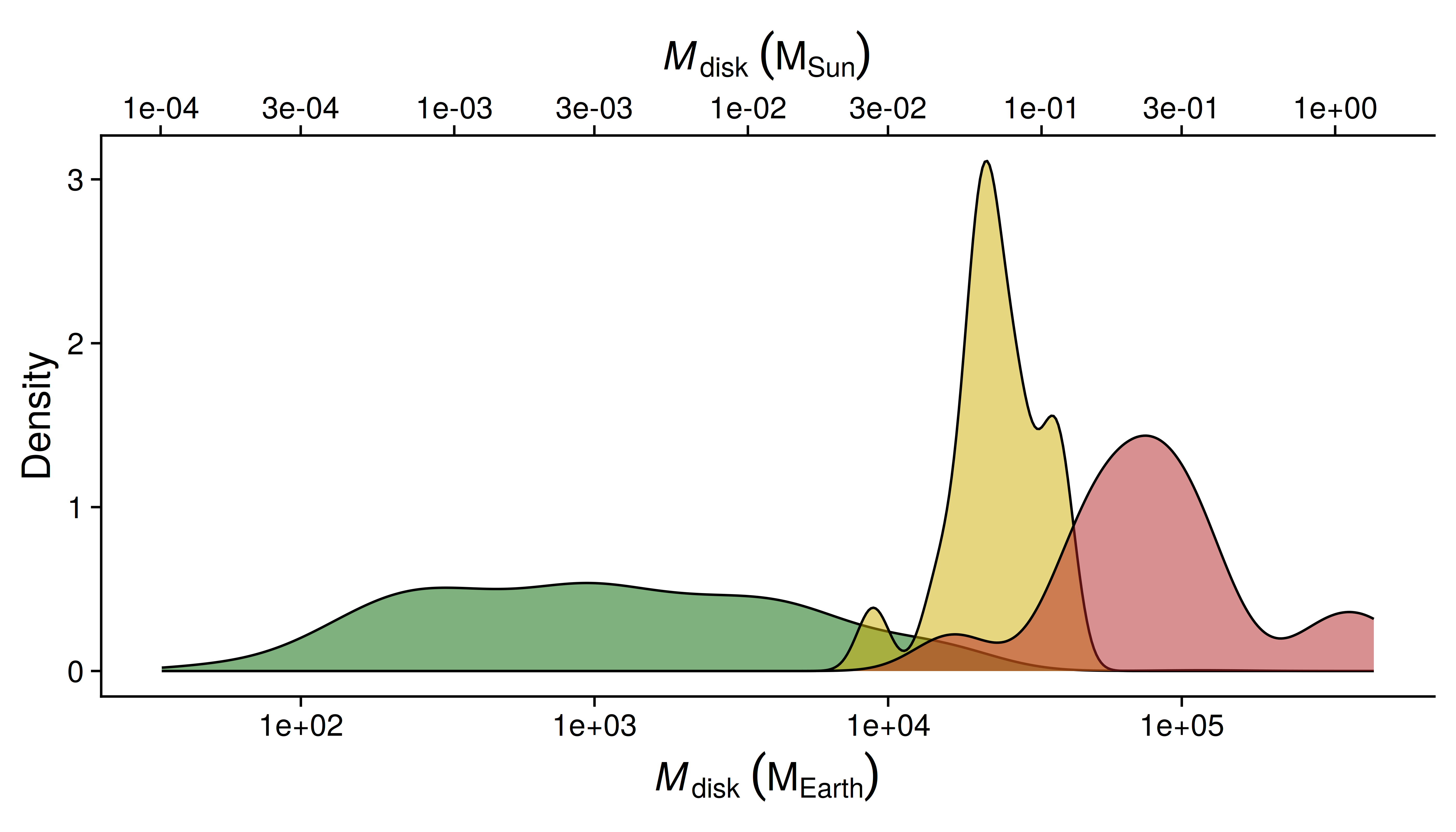}
    \caption{Density distribution of disk masses by stability type. Stable systems are shown in green, marginally stable in yellow, and unstable in red.}
    \label{fig:mass_density_by_stability}
\end{figure}

Figure~\ref{fig:mass_density_by_stability} presents the disk mass distributions for the three stability classes: gravitationally stable ($Q > 2$), marginally stable ($1 \leq Q \leq 2$), and gravitationally unstable ($Q < 1$). Analysis of these distributions reveals significant differences among the classes. Unstable disks exhibit substantially higher masses compared to their stable counterparts. Specifically, the median mass of stable disks is $3.4\times 10^{-3} M_\odot$, whereas for marginally stable and unstable systems, the median masses reach $6.8\times 10^{-2} M_\odot$ and $2.6\times 10^{-1} M_\odot$, respectively.

\begin{figure}[H]
    \centering
    \includegraphics[width=0.99\linewidth]{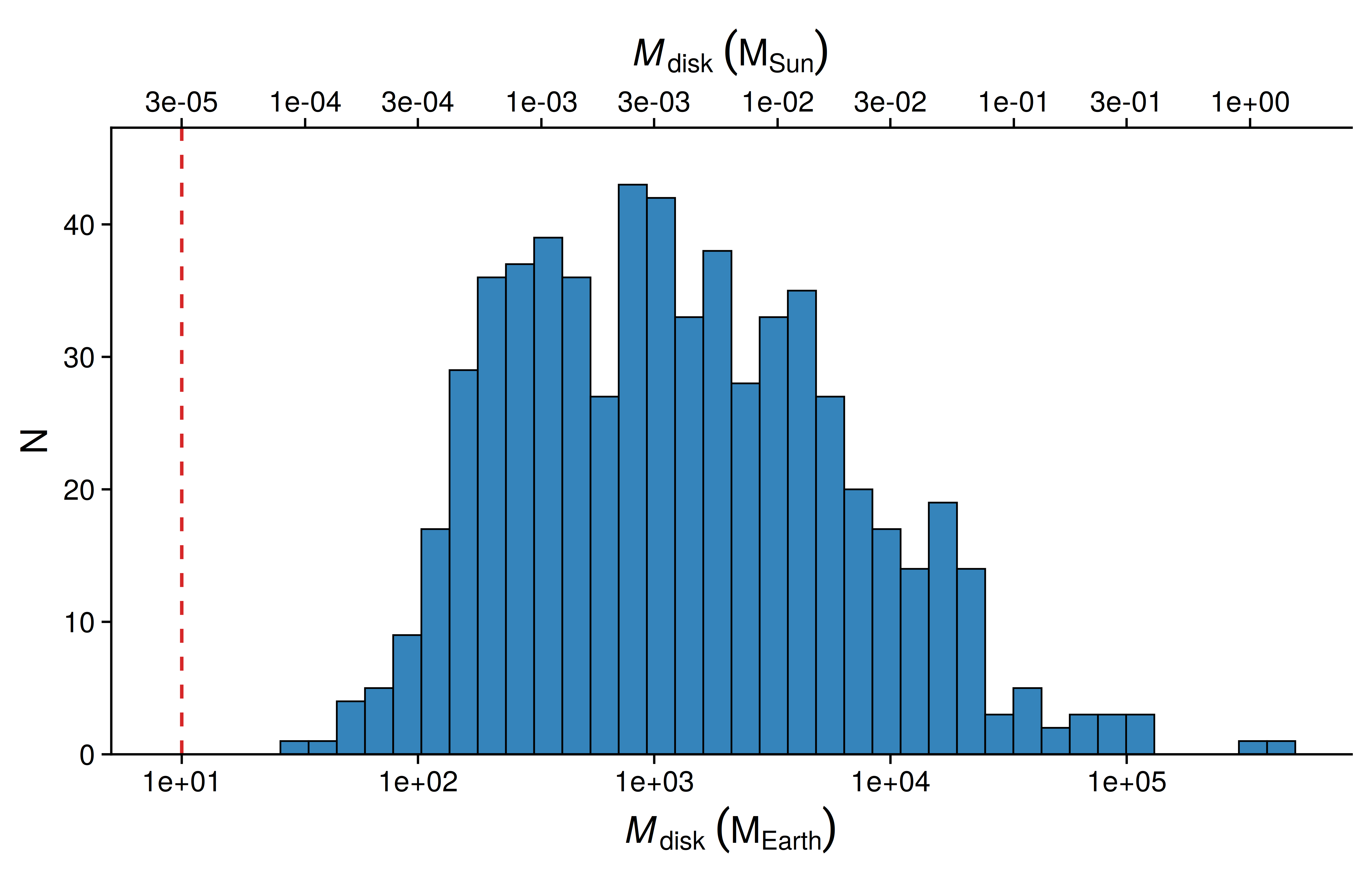}
    \caption{Overall mass distribution of the sample disks. The lower axis  shows masses in Earth masses, the upper axis in solar masses.  The vertical dashed line indicates the completeness threshold  of $10 M_\oplus$.}
    \label{fig:mass_distribution}
\end{figure}

\begin{figure}[H]
    \centering
    \includegraphics[width=0.99\linewidth]{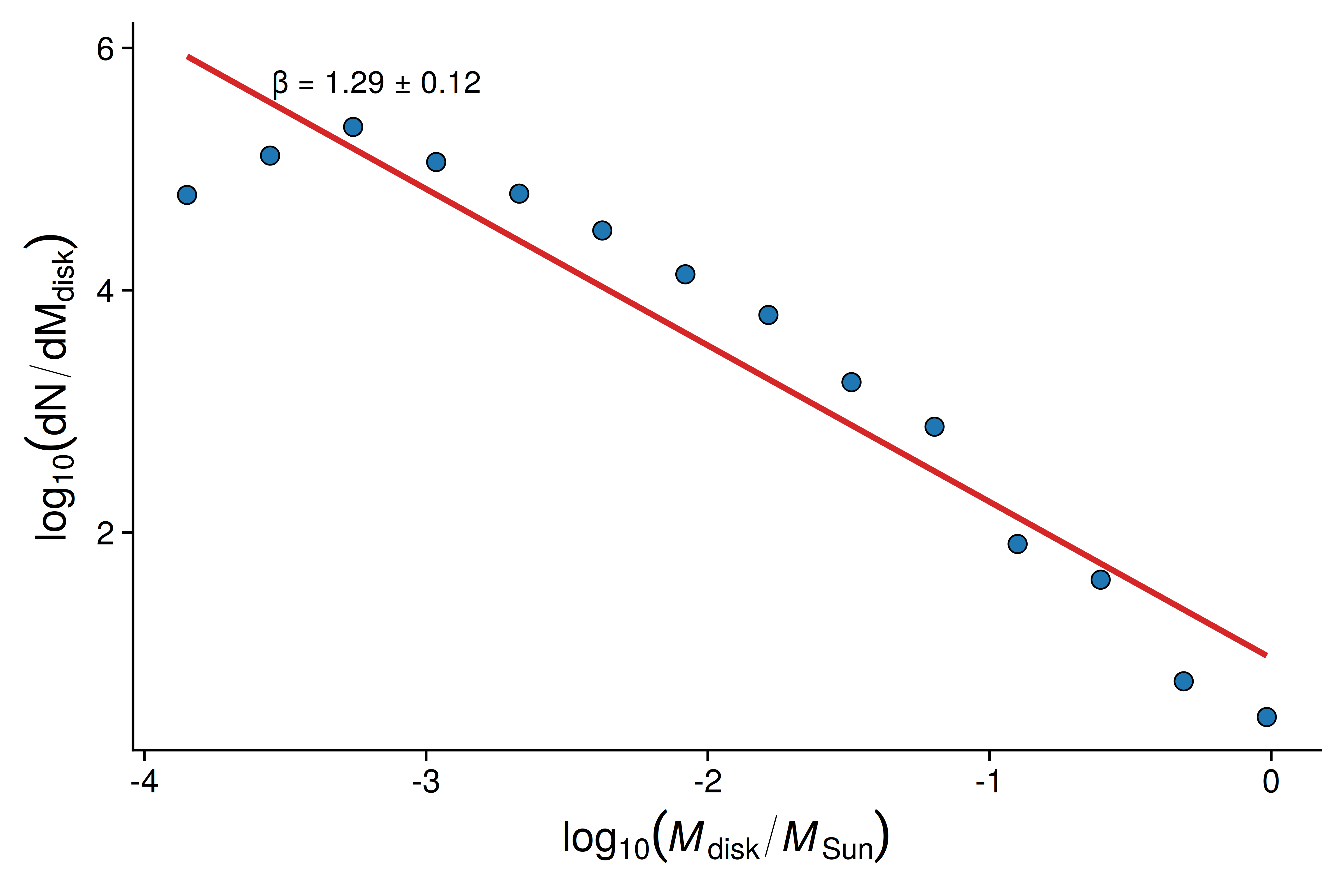}
    \caption{Power-law fit to the mass distribution of disks with 
             $M_{\text{disk}} > 10 M_\oplus$. The abscissa shows the 
             logarithm of mass in solar masses, the ordinate shows the 
             logarithm of the normalized distribution density $dN/dM$.}
    \label{fig:power_law_fit}
\end{figure}

Figure~\ref{fig:mass_distribution} shows the overall mass distribution of the sample disks. The distribution peaks at approximately $3\times 10^{-3} M_\odot$ and spans a range from $10^{-4}$ to $10^{-1} M_\odot$. The majority of systems cluster in the low-mass regime, while disks with masses on the order of $0.1$--$1 M_\odot$ represent only a small fraction of the sample. The vertical line in the figure indicates the completeness threshold ($10 M_\oplus$), which corresponds to the sensitivity limit of ALMA observations. For masses above this threshold, the distribution follows a power-law form (see Fig.~\ref{fig:power_law_fit}).

Figure~\ref{fig:power_law_fit} shows the power-law fit to the mass distribution for disks with masses $M_{\text{disk}} > 10 M_\oplus$, plotted in logarithmic coordinates. The points indicate the observed distribution density $dN/dM$, calculated as the number of disks per unit mass interval. The solid line represents the best-fit linear regression to the data. The resulting power-law index $\beta = 1.36 \pm 0.14$ yields a relation of the form $dN/dM \propto M^{-1.36}$. The coefficient of determination $R^2 = 0.885$ indicates that the power-law model accounts for $88.5\%$ of the variance in the data, demonstrating a good fit.

\begin{figure}[H]
    \centering
    \includegraphics[width=0.99\linewidth]{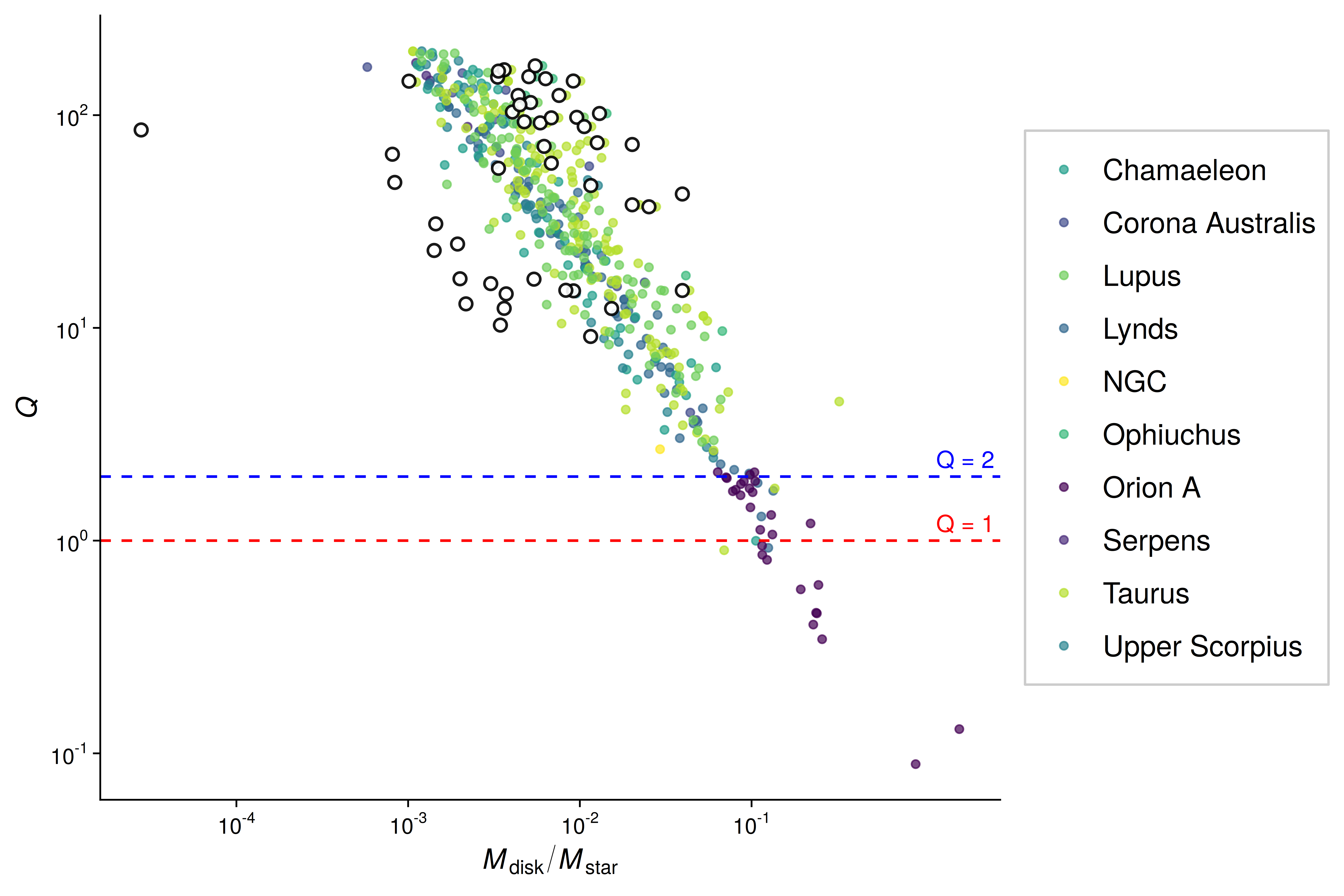}
    \caption{Dependence of the Toomre parameter $Q$ on the disk-to-star mass ratio. Dashed lines indicate the stability boundaries ($Q=1$) 
             and marginal stability ($Q=2$).}
    \label{fig:q_vs_mratio}
\end{figure}

Figure~\ref{fig:q_vs_mratio} illustrates the dependence of the Toomre parameter $Q$ on the disk-to-star mass ratio. As the mass ratio $M_{\text{disk}}/M_{\star}$ increases from $10^{-3}$ to $1$, $Q$ decreases from approximately $10^2$ to $10^{-1}$. When the mass ratio exceeds $0.132$, the $Q$ parameter falls below the critical value $Q = 1$, indicating the transition from gravitationally stable to unstable systems. This result aligns with theoretical predictions, which require $M_d/M_* \gtrsim 10^{-2}$ for the onset of gravitational instability ($Q < 1$) \citep{Kratter_2016}. The scatter in $Q$ values at a given mass ratio spans approximately an order of magnitude, reflecting variations in other system parameters such as temperature, radius, and surface density profile. Black open circles highlight 45 objects for which we have direct measurements of gas mass, dust mass, and stellar mass from the catalogs. For these objects, we estimate the Toomre parameter with the highest accuracy because we do not rely on additional assumptions about system parameters.

According to our analysis, $34$ objects ($2.9\%$) out of the $1155$ disks in the final sample exhibit signs of gravitational instability or lie at its boundary. Among these, we classify $14$ disks ($Q < 1$) as gravitationally unstable and $20$ disks ($1 \leq Q \leq 2$) as marginally stable. We provide a complete list of these systems in Appendix A (Table~\ref{tab:unstable_disks}).

\subsection{Comparison with Observed Disk Morphology}

Theoretical studies predict that the development of gravitational instability can lead to the formation of observable substructures in protoplanetary disks, including spiral arms, rings, and gaps \citep{Vorobyov2005}. To assess the consistency between the results of our statistical analysis of disk gravitational instability and the observed morphological characteristics, we compared the $Q$ values for objects in our sample with their ALMA images presented in the \texttt{Catalog of Circumstellar Disks}\footnote{\url{https://www.circumstellardisks.org/}}.

Table~\ref{tab:q_morphology} presents a comparison of the calculated Toomre parameter $Q$ values with the observed morphologies for protoplanetary disks resolved in ALMA observations. This comparison aims to investigate potential correlations between the stability parameter and specific morphological features, such as spiral arms, rings, and gaps. We obtained morphological data for a subset of objects from the DSHARP (Disk Substructures at High Angular Resolution Project) survey \citep{Andrews2018}. The DSHARP conducted ALMA observations at a frequency of $240$ GHz (wavelength $1.25$ mm) with an angular resolution of approximately $0.035``$, which corresponds to a linear resolution of $\sim5$ au at the typical distances of the star-forming regions in our sample. A key finding of the DSHARP project is that substructures are ubiquitous among bright protoplanetary disks, appearing in nearly all observed sources.

For six of the most well-studied disks in our sample—IM Lup, GM Aur, AS 209, HD 163296, MWC 480, and HL Tau—we performed a detailed comparison of our estimated $Q$ values with results reported in previous studies \citep{Sierra2021, Booth2020}. Table~\ref{tab:Q_key_disks} presents the Toomre parameter values we obtained, the observed morphological classifications for these disks, and the $Q$ estimates derived from multi-frequency ALMA observations and detailed modeling in the literature.

This comparison reveals a consistent association between the stability parameter and the observed morphology. Disks we classify as gravitationally stable ($Q > 2$) exhibit axisymmetric ring structures without evidence of global spiral perturbations. By contrast, disks with $Q \lesssim 1.7$, which lie near or below the threshold for gravitational instability, display distinct morphological signatures. Specifically, IM Lup shows prominent spiral structure, while MWC 480 and HL Tau occupy a borderline regime that requires more precise constraints on their gas masses to definitively assess their stability status.

These findings suggest that low $Q$ values ($\lesssim 1.7$) may serve as a reliable indicator of large-scale gravitational instability manifesting as spiral substructures in protoplanetary disks. The consistency between our statistical analysis and the detailed morphological characterization of individual systems supports the interpretation that the Toomre parameter captures the physical conditions conducive to instability, even in the presence of observational uncertainties.

\begin{strip}
\centering
\captionof{table}{Comparison of calculated Toomre parameter $Q$ with observed morphology}
\label{tab:q_morphology}
\begin{tabular}{llp{15cm}}
\hline
Object & $Q$ & Observed morphology \\
\hline
Sz 114   & 9.6 & Concentric substructures: narrow bright rings and dark gaps 
                  at distances ranging from several tens to more than 100 au 
                  from the star. \\
MWC 480  & 2.2 & Pronounced ring structure with narrow ($\lesssim10$ au) 
                  bright rings. \\
IP Tau   & 97.1 & Relatively smooth, axisymmetric surface brightness profile 
                  without prominent rings or gaps. \\
IQ Tau   & 13.7 & Presence of concentric features (rings and gaps). \\
Sz 129   & 11.0 & Complex system of concentric rings and gaps. \\
GW Lup   & 15.1 & Clearly defined ring structure. \\
IRC 101  & 0.78 & Pronounced ring structure. \\
V1094 Sco & 21.9 & Ring structure. \\
HT Lup A & 8.8 & Two-armed spiral structure. \\
\hline
\end{tabular}

    \vspace{0.3cm}
    
\centering
\captionof{table}{Comparison of $Q$ parameter: calculated values and estimates from other studies}
\label{tab:Q_key_disks}
\begin{tabular}{lllp{11.2cm}}
\hline
Object & $Q$ & Morphology & $Q$ estimates and stability status in other studies \\
\hline
IM Lup    & 1.3 & Spiral arms, ring & $Q < 1.7$, marginally stable \citep{Sierra2021} \\
MWC 480   & 2.2 & Rings and gaps    & $Q > 2$, stable (value close to threshold) \citep{Sierra2021} \\
GM Aur    & 4.2 & Rings, large cavity & $Q > 2$, stable \citep{Sierra2021} \\
AS 209    & 2.5 & Multiple narrow rings & $Q > 2$, stable \citep{Sierra2021} \\
HD 163296 & 4.9 & Several bright rings & $Q > 2$, stable \citep{Sierra2021} \\
HL Tau    & 1.8 & Concentric rings   & $Q < 1.7$ in the 50–110 au region, unstable \citep{Booth2020} \\
\hline
\end{tabular}
\end{strip}

Analysis of the data presented in Table~\ref{tab:q_morphology} and Table~\ref{tab:Q_key_disks} reveals a diversity of morphological types across the range of $Q$ parameter values. Disks with formally low $Q$ values near the stability boundary ($Q < 2$), such as MWC 480 ($Q = 2.2$) and IRC 101 ($Q = 0.78$), exhibit pronounced ring structures. However, as the comparison with well-studied objects demonstrates, confident classification of a disk as gravitationally unstable requires not only a formal $Q$ value below the critical threshold but also the presence of corresponding morphological signatures (e.g., spiral arms) and consistency with independent gas mass estimates.

Conversely, disks with high $Q$ values display a wide range of morphologies. These range from nearly smooth profiles, as seen in IP Tau ($Q = 97.1$), to complex systems of concentric rings and gaps, exemplified by Sz 114 ($Q = 9.6$), Sz 129 ($Q = 11.0$), and GW Lup ($Q = 15.1$). Notably, even spiral structures appear in disks with formally high $Q$ values, such as HT Lup A ($Q = 8.8$). The presence of spiral arms in such systems likely indicates local perturbations driven by mechanisms other than disk self-gravity, including interactions with a companion star or external environmental perturbations.

Taken together, this direct comparison demonstrates that the $Q$ value calculated from integral disk parameters does not uniquely predict the observed morphology, particularly for stable systems with $Q \gg 2$. Nevertheless, for disks near the stability boundary ($Q \lesssim 2$), low $Q$ values ($\lesssim 1.7$) appear to be more frequently associated with disks exhibiting signs of global gravitational instability manifesting as spiral arm structures. This suggests that while the Toomre parameter serves as a useful diagnostic for identifying gravitationally unstable systems, it should be interpreted in conjunction with morphological evidence and independent mass estimates.

\section*{Conclusion and Discussion}
\label{sect:outro}

In this work, we have performed a statistical analysis of the gravitational stability of protoplanetary disks using an extensive set of observational data. We compiled a sample of $1706$ objects by combining data from ten major surveys of star-forming regions. For each disk, we calculated the Toomre parameter $Q$ assuming a power-law surface density profile. Statistical processing of the derived $Q$ values and their comparison with the observed disk morphologies allowed us to estimate the prevalence of conditions favorable for the development of gravitational instability.

Our analysis reveals that only $1.2\%$ of the systems formally satisfy the criterion for gravitational instability ($Q < 1$). This finding is qualitatively consistent with the view that the phase of active gravitational instability represents a relatively short-lived evolutionary stage. However, as we discuss below, the observed fraction is one to two orders of magnitude lower than the predictions from numerical models.

Several factors may explain the low fraction of gravitationally unstable systems in our sample. First, the sample includes Class I and Class II objects, which correspond to later evolutionary stages characterized by small disk masses. Second, as our mass analysis demonstrates (Fig.~\ref{fig:mass_density_by_stability}), unstable disks exhibit high masses ($>10^4 M_\oplus$), making them intrinsically rare within the overall disk population.

The group of $20$ marginally stable disks ($1 \leq Q \leq 2$), which lie at the stability boundary, is of particular interest. These systems may represent a transitional stage in which gravitational instability is either just beginning to develop or, conversely, is decaying as a result of evolutionary mass loss.

Observational selection effects affect the sample we analyze, which we compiled from published catalogs of objects detected in the millimeter continuum with ALMA \citep{Williams2011}. The sample primarily includes disks with the highest flux, that is, the most massive disks and those with high dust surface density. This selection biases the sample toward more massive systems, particularly for distant regions such as Orion ($\sim400$~pc). This bias, combined with the distinct physical conditions in this massive star-forming region \citep{Testi2022}, explains the concentration of disks with low Toomre parameter values ($Q < 2$) in the Orion region (29 out of 34 objects; see Appendix). We must account for this selection effect when interpreting the absolute fractions of unstable disks; however, it does not invalidate relative comparisons between subsamples or the identification of general statistical trends, which also appear in other studies \citep{Sierra2021}.

Our comparison of the calculated Toomre parameters with the observed disk morphologies (Table~\ref{tab:q_morphology}) reveals interesting patterns. Disks with low $Q$ values, such as IRC 101 ($Q = 0.78$) and MWC 480 ($Q = 2.2$), exhibit complex ring structures, which may potentially arise from active gravitational processes in these systems. Notably, in the HL Tau disk, where CO isotopologue line emission also indicates possible gravitational instability \citep{Booth2020}, we find a low $Q$ value ($Q < 2$), consistent with this interpretation.

Our results reveal a discrepancy with predictions from numerical simulations. Modern gas-dynamic calculations of long-term system evolution demonstrate that massive disks can remain in a phase of active fragmentation and episodic accretion for several hundred thousand to a million years, constituting a significant fraction of their lifetime \citep{Matsukoba2023}. If we assume that the observed fraction of disks with $Q < 2$ ($\sim 2.9\%$) reflects the proportion of systems currently in this phase, then the characteristic duration of such a phase would be only $\sim 12$ thousand years. This estimate is one to two orders of magnitude shorter than model predictions.

This contradiction points to two possible scenarios. In the first scenario, the observed substructures (spirals, rings, clumps) arise from mechanisms other than gravitational instability. In the second and more likely scenario, systematic underestimation of disk masses in observational data leads to an overestimation of the Toomre parameter $Q$ and, consequently, to a systematically underestimated fraction of gravitationally unstable systems. Studies demonstrating that current methods may underestimate actual disk masses by an order of magnitude or more support this hypothesis \citep{Miotello2017, Williams2011}.

We must consider our results in the context of methodological limitations inherent in studies of protoplanetary disk gravitational stability. The Toomre parameter characterizes the local stability of the disk and is sensitive to radial variations in physical conditions \citep{Kratter_2016}. Investigating global instability modes requires more sophisticated criteria that account for the full disk structure.

The main sources of uncertainty in our $Q$ estimates include:

\begin{enumerate}
    \item \textbf{Systematic underestimation of masses}. The use of the standard gas-to-dust ratio of $100$ and the assumption of optically thin emission may lead to underestimation of actual disk masses. According to estimates by \citet{Miotello2017}, actual masses may be underestimated by an order of magnitude or more. This estimate aligns with independent assessments of individual error sources: dust optical depth effects can lead to mass underestimation by a factor of $\sim5$--$30$ \citep{Semenov2003}; chemical depletion of CO contributes a factor of $\sim2$--$10$ \citep{Miotello2017}; uncertainty in the gas-to-dust ratio adds a factor of $\sim2$--$3$; and additional uncertainties relate to dust temperature and particle size distribution \citep{Agladze1996}. An important additional factor is the growth of dust to millimeter sizes in early disks, which leads to a sharp increase in opacity and, consequently, to significant underestimation of dust mass estimates in ALMA Band 6 observations, according to modern models \citep{Vorobyov2026}.
    
    \item \textbf{Influence of magnetic fields}. Magnetic fields exert a complex influence on disk gravitational stability, reflected in modified stability criteria \citep{Dudorov2015}. On one hand, magnetic pressure and tension have a stabilizing effect; on the other hand, magnetic forces can reduce the effective rotation velocity, making the disk more prone to instability \citep{Lizano2010}. In protoplanetary disks, the stabilizing effect typically dominates, suppressing fragmentation and planet formation via gravitational instability \citep{Lizano2010}. Numerical simulations incorporating the joint evolution of magnetorotational and gravitational instabilities show that magnetic fields can also lead to episodic accretion and the formation of dust rings in the inner disk regions \citep{Vorobyov2020}. However, direct measurement of magnetic fields in protoplanetary disks remains extremely challenging, and reliable Zeeman measurements exist for only a few objects \citep{Khaibrakhmanov2024}.
    
    \item \textbf{Local variations in parameters}. Estimating $Q$ from averaged parameters may not reflect conditions in specific regions where prerequisites for instability may develop. Accurate diagnostics require radial maps of the Toomre parameter distribution across the disk.
    
    \item \textbf{Uncertainties in size determination}. Accurately determining the linear radii of disks remains one of the most challenging problems in observational astrophysics. Typically, we can measure only the radius of the dust component, while the gas disk may be two to three times larger \citep{Grant2021}. This significantly affects surface density estimates and, consequently, the Toomre parameter. Theoretical models of disk evolution show that the initial radii of gas disks should be $\sim100$ au to match the observed distributions in regions of different densities \citep{Marchington2022}.
    
    \item \textbf{Uncertainties in stellar mass determination}. Accurately estimating central star masses requires a comprehensive analysis of their position on the Hertzsprung--Russell diagram, which presents significant difficulties when working with large statistical samples.
\end{enumerate}

The cumulative effect of the aforementioned factors, which leads to potential underestimation of disk masses and correspondingly incorrect conclusions about their gravitational stability, suggests that the actual fraction of systems satisfying the gravitational instability criterion ($Q < 1$) may be substantially higher than we observe.

Our choice of a specific power-law density profile does not affect the overall conclusion regarding disk gravitational stability within the framework of our adopted approach. Although the radial density distribution may determine the location of potential gravitationally unstable regions within the disk \citep{Vorobyov2020}, the very development of gravitational instability depends primarily on the total disk mass and its ratio to the stellar mass. In particular, a profile with an index of $-1.5$, obtained in numerical simulations of accretion bursts in protoplanetary disks \citep{Vorobyov2020}, and the profile with an index of $-1$ that we use in this work lead to qualitatively similar conclusions about the overall stability of the system.

To overcome these limitations and obtain more accurate estimates of gravitational stability, we must apply methods less sensitive to optical depth, including observations of rare CO isotopologues and the HD molecule. We should also transition from averaged estimates to radial maps of the Toomre parameter distribution based on high-resolution data and conduct targeted studies of Class 0 and Class I disks, where conditions for the development of gravitational instability are most favorable. Improving telescope resolution to accurately determine dust disk radii, followed by extrapolation to the gas component, also represents an important direction for future work.

Our statistical analysis allows us to draw the following key conclusions:

\begin{enumerate}
    \item Only $1.2\%$ of the disks in our sample formally satisfy the gravitational instability criterion ($Q < 1$), and $1.7\%$ reside in a state of marginal stability ($1 \leq Q \leq 2$).
    
    \item A significant discrepancy exists between the observed fraction of unstable systems and the predictions of theoretical models. This discrepancy indirectly supports the hypothesis that current observations substantially underestimate protoplanetary disk masses.
    
    \item Our analysis of objects with resolved morphology reveals that systems with $Q < 2$ often exhibit complex ring or spiral structures, consistent with predictions from gravitational instability models.
\end{enumerate}

Thus, the statistical analysis methodology we propose enables the preliminary identification of systems potentially susceptible to gravitational instability and can serve as a starting point for more detailed studies of individual objects.

\begin{acknowledgements}
The authors thank V. V. Akimkin for valuable comments that helped significantly improve 
this work, as well as the referee for constructive recommendations and 
useful advice. The work of S. A. Khaibrakhmanov was supported by the 
Theoretical Physics and Mathematics Advancement Foundation `BASIS`
(project 23-1-3-57-1).
\end{acknowledgements}

\bibliographystyle{aa}
\bibliography{references}

\onecolumn 
\begin{appendix}

\section{List of Disks with Signs of Gravitational Instability}
\label{appendix:unstable_disks}

Table~\ref{tab:unstable_disks} presents a list of 34 disks whose calculated 
parameters show signs of gravitational instability ($Q < 1$) or correspond 
to a state of marginal stability ($1 \leq Q \leq 2$).

\begin{table}[H]
\centering
\caption{List of disks with signs of gravitational instability}
\label{tab:unstable_disks}
\begin{tabular}{llrrr|llrrr}
\hline
\multicolumn{5}{c|}{\textbf{Unstable disks ($Q < 1$)}} & 
\multicolumn{5}{c}{\textbf{Marginally stable disks ($1 \leq Q \leq 2$)}} \\
\hline
Object & Region & $M_{\rm d}$ & $M_*$ & $Q$ & 
Object & Region & $M_{\rm d}$ & $M_*$ & $Q$ \\
 &  & [$M_\odot$] & [$M_\odot$] &  &  &  & [$M_\odot$] & [$M_\odot$] &  \\
\hline
FRM2016 185 & ONC & 0.308 & 0.57 & 0.13 & MGM129 & Ori A & 0.037 & 0.85 & 1.32 \\
MLLA 642 & ONC & 0.451 & 1.50 & 0.08 & MGM371 & Ori A & 0.028 & 1.10 & 1.70 \\
MLLA 381 & ONC & 0.116 & 1.35 & 0.35 & MGM399 & L1641 & 0.019 & 0.50 & 1.30 \\
MLLA 404 & ONC & 0.099 & 1.30 & 0.41 & MGM612 & Ori A & 0.028 & 1.04 & 1.74 \\
MLLA 440 & ONC & 0.087 & 1.10 & 0.46 & MGM512 & L1641 & 0.009 & 0.20 & 1.72 \\
MLLA 424 & ONC & 0.088 & 1.10 & 0.46 & 04142626+2806032 & Tau & 0.017 & 0.37 & 1.76 \\
MLLA 385 & ONC & 0.068 & 1.05 & 0.59 & 05351470-0522396 & ONC & 0.028 & 0.85 & 1.43 \\
MLLA 422 & ONC & 0.064 & 0.79 & 0.62 & FRM2016 208 & ONC & 0.021 & 0.70 & 1.90 \\
FRM2016 206 & ONC & 0.049 & 1.20 & 0.82 & FRM2016 224 & ONC & 0.024 & 0.70 & 1.69 \\
MLLA 401B & ONC & 0.047 & 1.21 & 0.85 & MLLA 476 & ONC & 0.022 & 0.75 & 1.85 \\
04215740+2826355 & Tau & 0.126 & 5.47 & 0.90 & MLLA 426 & ONC & 0.020 & 0.85 & 1.97 \\
MGM378 & L1641 & 0.017 & 0.40 & 0.92 & MLLA 572A & ONC & 0.021 & 0.60 & 1.91 \\
MLLA 389 & ONC & 0.042 & 1.10 & 0.95 & FRM2016 191 & ONC & 0.020 & 0.85 & 1.98 \\
 & & & & & 05351346-0522105 & ONC & 0.023 & 0.70 & 1.76 \\
 & & & & & MGM307 & L1641 & 0.014 & 0.40 & 1.87 \\
 & & & & & MGM269 & L1641 & 0.011 & 0.35 & 2.06 \\
 & & & & & MLLA 579 & ONC & 0.024 & 0.85 & 1.63 \\
 & & & & & MGM762 & L1641 & 0.011 & 0.40 & 2.15 \\
 & & & & & MGM540 & Ori A & 0.040 & 0.55 & 2.00 \\
 & & & & & 05351474-0522335 & ONC & 0.019 & 0.55 & 2.10 \\
\hline
\end{tabular}
\end{table}

\end{appendix}

\end{document}